\documentclass[preprint]{aastex}

\usepackage{graphics}
\DeclareGraphicsExtensions{.eps}

\begin{document}
\title{The ROTSE-III Robotic Telescope System}

\author{C.~W.~Akerlof, R.~L.~Kehoe,\altaffilmark{1} T.~A.~McKay, E.~S.~Rykoff and
  D.~A.~Smith}
\affil{University of Michigan, Department of Physics, 2477 Randall
Laboratory, Ann Arbor, MI 48109-1120 USA; cakerlof@umich.edu, kehoe@pa.msu.edu,
  tamckay@umich.edu, erykoff@umich.edu, donaldas@umich.edu}

\author{D.~E.~Casperson, K.~E.~McGowan, W.~T.~Vestrand, P.~R.~Wozniak and J.~A.~Wren}
\affil{Los Alamos National Laboratory, NIS-2   MS D436, Los Alamos NM
87545 USA; dcasperson@lanl.gov, mcgowan@aslan.lanl.gov, vestrand@lanl.gov,
wozniak@algol.lanl.gov, jwren@nis.lanl.gov}

\author{M.~C.~B.~Ashley and M.~A.~Phillips}
\affil{School of Physics, The University of New South Wales, Dept. of
Astrophysics and Optics, Sydney 2052, Australia; mcba@phys.unsw.edu.au,
a.phillips@unsw.edu.au}

\author{S.~L.~Marshall}
\affil{Lawrence Livermore National Laboratory, University of
California, P.O. Box 808, Livermore, CA 94551-0808 USA; stuart@igpp.ucllnl.org}

\author{H.~W.~Epps}
\affil{Lick Observatory, University of California Observatories, Santa
Cruz, CA 95064 USA; epps@ucolick.org}

\and

\author{J.~A.~Schier}
\affil{The Pilot Group, 128 West Walnut Avenue, Unit C, Monrovia, CA
91016 USA; alan@the-pilot-group.com}

\altaffiltext{1}{Also at Michigan State University, 208 Physics-Astronomy
  Bldg., East Lansing, MI 48824-1116 USA}

\begin{abstract}
The observation of a prompt optical flash from GRB990123 convincingly demonstrated the value of autonomous robotic telescope systems. Pursuing a program of rapid follow-up observations of gamma-ray bursts, the Robotic Optical Transient Search Experiment (ROTSE) has developed a next-generation instrument, ROTSE-III, that will continue the search for fast optical transients.  The entire system was designed as an economical robotic facility to be installed at remote sites throughout the world. There are seven major system components: optics, optical tube assembly, CCD camera, telescope mount, enclosure, environmental sensing \& protection and data acquisition. Each is described in turn in the hope that the techniques developed here will be useful in similar contexts elsewhere.

\end{abstract}

\section{Introduction}

For many years, it was realized that progress in understanding gamma-ray bursts (GRBs) would depend on accurate localizations only attainable at longer wavelengths. Following the premature failure of the onboard tape recorders on the Compton Gamma-Ray Observatory in 1992, the shift to real time data transmission opened the window for ground-based fast response systems that could slew to designated targets within ten seconds. The first optical system to exploit this capability was the GROCSE-I wide-field camera at Lawrence Livermore National Laboratory \citep{lee97, park97a}. The hardware was originally conceived under the aegis of the Strategic Defense Initiative as a test bed for developing the technologies to rapidly locate hostile incoming missiles and launch appropriate countermeasures. By the time the GROCSE collaboration obtained the use of this instrument in 1993, it had been abandoned and neglected for several years. The collaboration invested considerable effort in adapting this device to respond robotically to the GRB trigger messages relayed from the Goddard Space Flight Center alert system called BACODINE ({\bf BA}TSE {\bf CO}ordinates {\bf DI}stribution {\bf NE}twork), later renamed GCN \citep{sdb97} ({\bf G}RB {\bf C}oordinates {\bf N}etwork).

The sensitivity of this device was limited to $m_v \sim 9$ by the restricted angular acceptance of the fiber optic light guides that coupled the spherical focal surface to the array of image intensifiers and the modest quantum efficiency of the intensifier photocathodes. Probably no more than 0.3\% of all photons entering the 62 mm aperture lens were available to contribute to the detected image. It was soon realized that a much more powerful system could be constructed with fast 35 mm camera lenses and large format CCD image sensors. This much simpler system promised at least 5 magnitudes greater sensitivity and a field of view to match BATSE error boxes, typically $16^\circ \times 16^\circ$, yet could still be built for $\sim$ \$200,000. This overall design was implemented within a few years as two competing systems, LOTIS at Lawrence Livermore National Laboratory in California and ROTSE-I at Los Alamos National Laboratory in New Mexico. This design choice proved prescient when, on January 23, 1999, the ROTSE-I camera array recorded the first contemporaneous optical flash from an exceptionally bright gamma-ray burst \citep{akerlof99}. However, in several years of operation, no other similar events were detected by either ROTSE \citep{akerlof00, kehoe01} or LOTIS \citep{park97b, williams99}. Clearly, a rapid response instrument was needed with a limiting sensitivity exceeding ROTSE-I/LOTIS but not necessarily better than $m_v \sim 20$ that is exhibited by typical GRB optical afterglows many hours after the original event.

In fact, several years before the first GRB afterglow was detected on February 28, 1997 \citep{groot97}, Akerlof proposed building two 0.45-meter aperture wide-field telescopes to systematically search BATSE error boxes in an hour or so following a GRB detection in space. With support from the Research Foundation and NASA, the telescopes, called ROTSE-II, were designed and built. Unfortunately, a variety of flaws compromised the ability of these instruments to reach the original objectives.

The detection of a contemporaneous optical flash from GRB990123 prompted a
thorough re-evaluation of our ROTSE-II development program. Three design
requirements received special attention: (1) a fast slewing telescope mount;
(2) a CCD camera with a high quantum efficiency image sensor; (3) a simpler
optical design. Furthermore, if system costs could be contained, we might be
able to build a sufficient number of identical instruments to distribute an
effective global array of observatories. Six instruments, three in each
hemisphere, would be sufficient to promptly observe a large fraction of all
GRBs, 24 hours per day. The logistical problem of installing a complete,
functioning instrument anywhere in the world forced a modular approach to the
overall design. Taking a lesson from the GONG helioseismology project, we
required that the telescope enclosure dimensions permit shipment via a 20'
standard seagoing container unit. This constrained the width and height of the
telescope enclosure to slip through the $92" \times 90"$ container door
frame. We selected a vertical cylinder 90" in diameter and 86.25" high to keep
the structure simple and rugged. The telescope tube assembly must rotate freely
within the enclosure roof opening. With a square roof aperture 58" on a side,
the optical tube assembly was limited to a maximum swing radius of 29 inches. This put a stringent limit on permissible optical designs.

The original optical design for ROTSE-II was predicated on the need to search a
$16^\circ \times 16^\circ$ error box within an hour, demanding a field of view
as large as possible. With a greater emphasis on GRB follow-up observations
from missions like HETE-2 and Swift that promised much higher accuracy
localizations, this requirement was no longer dominant. However, we had learned
from ROTSE-I the value of continual repetitive observations of the night
sky. In this case, our secondary goal was a search for `orphan' optical
transients that might signal GRB-like events whose energetic gamma-ray beams
happened not to intersect the Earth. An orphan search with ROTSE-I data
indicated a wide-field instrument would be ideal to search for these transients
{\citep{kehoe02}.
Thus, we opted to continue our efforts to
build an f/1.9, 0.45-meter aperture instrument despite the realization that we
would sacrifice about a factor of two in signal-to-noise ratio. These
parameters were not entirely selected out of thin air: we were acquainted with
the US Air Force satellite tracking telescope systems called GEODSS. These are
1.0-meter aperture telescopes \citep{jeas81, beatty82}, operating with a focal
ratio of f/2.1. Designed in the late `70s by David Grey, these devices have
played a major role in detecting Earth-crossing asteroids \citep{pravdo99,
stokes00}. 

The prompt detection of GRB990123 was a landmark for robotic telescopes: it is probably the first significant scientific result for which the autonomous robotic performance was absolutely critical to success. We have tried to incorporate essentially all of the useful features of ROTSE-I into the present ROTSE-III system. It is hoped that some of these ideas may be useful to others who are interested in building low maintenance autonomous remote observatories.

\section{Optical Design}

The optical design was strongly constrained by the demand for a compact instrument with a swing radius of not more than 29", which could be well baffled against stray light and which would be relatively insensitive to decollimation issues.  A high premium was placed on simplicity and upon anticipated ease of optical manufacture as the ROTSE-III program called for multiple copies to be constructed in a timely manner and at an affordable cost.

Initial specifications for the telescope/camera outlined a (modified)
Cassegrain with a 450.0-mm diameter f/1.80 primary mirror and an all-refracting
field corrector providing a final focal length of 850.0 mm, with the final
focus located not more than 75.0 mm in front of the primary-mirror vertex.  A
(0.40 to 0.90)-micron passband was chosen and a $2{^\circ}.64$ diameter flat
field of view (f.o.v.) was stipulated so as to cover a 2048$\times$2048 Marconi
CCD with
13.5~micron pixels to its corners.  The pixel scale (3.28 arcsec/pixel) will undersample the ``seeing" but that limitation was accepted in order to attain the desired $2{^\circ}.64$ field coverage.

The specifications also called for the inclusion of a flat broad-passband
colored-glass filter and space for a Prontor magnetic E/64 shutter within the
field corrector, as well as a flat fused silica vacuum Dewar window and a back
focal distance (b.f.d.) not less than 7.7 mm.  The image quality goal was to
enclose 70\% or more of the incident energy inside a 13.5 micron diameter at all field angles and wavelengths without refocus, with not more than 7 microns of maximum rms lateral color over the full spectral range.

Modeling and optical optimization were done (by HWE) with his proprietary code, OARSA.  Some of the figures and the subsequent image analysis presented here were computed with the commercially available code, ZEMAX$^\copyright$ which provides a convenient independent assessment of the system.

An initial exploration of the design demonstrated that a refracting corrector
composed of 4 free-standing powered lens elements made of the same material
would satisfy the requirements.  Similar 3-element correctors were first
suggested by R.A.~Sampson \citep{sampson13a, sampson13b}
and pioneered algebraically by C.G. Wynne \citep{wynne49}.  Aberration control as well as remarkable color correction are achieved by adjusting the power ratios among the lens elements.  An early practical example of a 4-element corrector of this type is one designed at f/3.52 for the Palomar 5-m prime focus \citep{wynne67}.

Preliminary optimizations demonstrated that hyperbolic (or higher-order aspheric) correction on the primary mirror would not be needed so a parabolic primary mirror was chosen for relative ease of manufacture and testing.  Further calculations showed that the extra degrees of freedom provided by a curved secondary mirror would not be required.  Thus the system became a prime focus design in effect.  A flat secondary mirror was retained to fold the optical system so as to meet mechanical constraints.  An overview of the optical design is shown in Figure~\ref{fig:one}.

Subsequent optimizations were carried out to explore possible optical glass alternatives for the 4-element all-spherical corrector.  It was found that a variety of crown glasses could be used for the first 3 lens elements, as long as all of those glasses were the same in a given design.  However it proved advantageous to use a lower-index, lower-dispersion glass type for the last lens element.  Ohara S-FPL51Y (497811) was found to be the best choice.  Ohara BAL35Y (589612) was chosen for the leading lens elements in the construction design, based upon a combination of factors including ready availability, excellent internal transmission, reasonable cost in the required (relatively small) sizes and best image quality.  A flat 4.0-mm-thick piece of Schott BK7 (517642) was used as the colored-glass filter simulator.

Eight identical sets of optical glass were ordered from Ohara Corporation, based upon a preconstruction version of the optical design.  High precision (6-place) melt-sheet indices of refraction for the BAL35Y and S-FPL51Y glass types were measured by Ohara at each of 10 standard emission-line wavelengths in the (0.365 to 1.014)-micron spectral range.  Measured melt indices were transformed to the nominal system operating temperature, T=+10.0$^\circ$~C, using published thermo-optical constants (dn/dT values) from the Ohara catalog.  The transformed indices were then fitted differentially to the dispersion curves of the respective generic glass types.  Indices calculated from these fits were subsequently fitted to the Schott formula.  The resulting Schott dispersion constants are given for each of the melt glass types in Table~\ref{tab:one} which also includes catalog Schott BK7 and fused silica at T=+10.0$^\circ$~C.

Tucson Optical Research Corporation (TORC) was identified as the optical manufacturer for the refractive components and the construction design was optimized using standard TORC test-plate radii for all of the corrector-lens surfaces.  Construction design Run No. 061900AC is the best in a sequence of optimizations of this general description.  Its system prescription is given in full quantitative detail at T=+10.0$^\circ$~C in Table~\ref{tab:two}.

This construction optical design shows rms image diameters of 6.6 $\pm$ 2.2 microns averaged over all field angles and wavelengths within the (0.40 to 0.90)-micron passband without refocus, with 1.9 microns of maximum rms lateral color.  These image characteristics, calculated in real time during the design optimization process, suggest that the expected image quality is better than the specified requirements.  That is confirmed by Figure~\ref{fig:two} which shows the polychromatic fractional encircled energy as a function of image radius, at 5 field angles which are:  on-axis, 39\%, 78\%, 96\% and 100\% of the full CCD f.o.v.  Each of the 5 polychromatic ray traces is composed of equal numbers of randomly distributed rays at (0.40, 0.43, 0.47, 0.52, 0.58, 0.65, 0.73, 0.81 and 0.90)-micron wavelengths.

It is evident in Figure~\ref{fig:two} that the residual aberrations have been distributed so as to produce a strong gradient toward the outer 10\% of the field radius, where there is very little detector area available (in the corners).  Nevertheless, it can be seen that some 72\% of the energy lies within a 13.5-micron diameter for the worst-case image at full field while the other images are considerably tighter.

\section{Mechanical Design and Fabrication}

The parabolic primary mirrors for ROTSE-III have been figured by Don Loomis of
Tucson; the refractive components and secondary flat were produced by Tucson
Optical Research Corporation (TORC). Although the CCD pixel size of 13.5
microns relaxes the optical tolerances so that diffraction-limited performance
is not required, it was still necessary to hold the primary surface to ${1\over
4}$-wave peak-to-peak and ${1\over 20}$-wave smoothness to maintain a
satisfactory point spread function. The refractive components were
anti-reflection coated by TORC and the mirrors were silver coated by Newport
Thin Films and Denton Vacuum. J.~Alan Schier designed the optical tube
assembly. Considerable effort was exerted to minimize the weight and moment of
inertia that sets the scale for the telescope slewing torques. The corrector
cell lenses were mounted with Delrin spacers to provide radial thermal
compensation. The primary mirror is also athermally supported with radial
Delrin spacers and longitudinally constrained at three points, 120$^\circ$ apart. There was some concern that this would lead to surface deflection under gravity that would degrade stellar images. Dr.~Anna Moore assisted us by running finite element calculations of these effects, showing that such distortions would not be a serious problem.

No attempt was made to compensate for longitudinal temperature expansion. The focal plane temperature shift was calculated to correspond to a secondary mirror motion of 7.9 microns/$^\circ$C, in excellent agreement with experimental measurements. Since the total operating thermal environmental range is less than 50$^\circ$~C, the maximum secondary mirror travel is less than 0.5 mm. This permitted a simple but rugged design for controlling the mirror position and providing focus adjustment. The secondary mirror cell is suspended by two thin stainless steel annuli, perforated with diagonal slots. These sheet metal rings are flexible enough to permit a 1.0 mm longitudinal travel while rigidly constraining the transverse motion and prohibiting tilt. A small motorized micrometer assembly drives the focal motion against a spring-loaded restoring force. The focus distance does shift slightly with elevation angle as well as temperature but no azimuthal dependence is apparent. This has been empirically modeled with a simple mathematical function, yielding a residual RMS error of 10 microns. No hysteresis has been observed, evidence of a good rigid structure.

\section{CCD Camera}

The CCD camera design was predicated on the use of the Marconi Applied Technologies CCD42-40-2-343 back-illuminated 2048 $\times$ 2048 sensor with 13.5 $\mu$m pixels. This device was selected on the basis of its high quantum efficiency (QE) and low readout noise at 1 MHz clock rates. The increased QE promised a factor of two improvement in sensitivity, equivalent to increasing the telescope aperture by a similar factor. Astronomical Research Cameras was selected to build the mechanical housing and the readout electronics. The original camera was designed with an air-cooled Peltier cooling system to avoid handling any kind of cryogenic or room temperature fluids. This led to some engineering problems that favored a more compact liquid-cooled device. The current camera relies on a propylene glycol heat transfer loop driven by a Polyscience model 340 air-cooled recirculator. In operation, the camera can routinely reach -40$^\circ$~C in an ambient environment of +20$^\circ$~C.  To allow fast readouts without bus contention problems, the camera interface is mounted in a dedicated PC running under the Linux OS.  The image files are stored on NFS cross-mounted disks resident on the data acquisition computer and commands are communicated via TCP/IP socket protocols.

The CCD42-40 chip can be read from two independent amplifiers or in single-ended mode at half the speed. This choice is software selectable. We have been using the camera principally in the latter mode with a total readout time of 6 seconds. A drawback to dual-channel operation is the presence of cross-talk from bright star images. For studying the rapid variability of GRBs at the earliest times following a burst, this deficiency is less onerous. The readout noise of the current camera is expected to be $\sim 10~{\rm e}^-$ while the sky noise is of the order of 15 e$^-$ per second. Thus, the image noise is sky dominated within 10 seconds. Exposures of 5 s, 20 s and 60 s reach limiting magnitudes of 17, 17.5 and 18.5 at the test site at Los Alamos National Laboratory. This is expected to improve at the darker sites where these systems will be permanently located.

\section{Telescope Mount}

The scarcest component of the ROTSE-III telescope system was the equatorial fork mount. With an 18" aperture, the optical tube assembly was just beyond the capacity of telescope mounts for the amateur mass market and our scientific requirements for fast slew and accurate tracking further aggravated the problem. Such systems are frequently built for military customers, but with price tags starting at \$100,000. We were extremely fortunate that just as this problem became acute, the Astro Works Corporation introduced a modestly priced instrument for deep sky photography called the ``Centurion 18". The telescope fork was constructed from stamped and welded sheet steel, providing an optimal strength-to-weight ratio, making this an ideal starting point for a low-inertia, rapid slew platform.

These mounts were subsequently modified by J.~Alan Schier to provide higher slewing torques, more accurate tracking and improved mechanical tolerances. The original stepper motors were replaced by Pittman servomotors in friction contact with 4" diameter hardened steel disks fixed to the right ascension and declination axes. The angular position is derived from incremental encoders consisting of precision ruled tapes with 10 micron spacing in close proximity to opto-electronic detectors that sense relative motion. The absolute angular location is derived from counting the ruling marks relative to mechanical home sensor switches. One problem with a welded steel frame is that accurate co-alignment of the declination axis bearings is difficult to obtain. This potential pitfall was evaded by rigidly constraining the drive-side shaft while thinning the shaft on the opposite fork to allow a small amount of flexure.

The mount control code is written in C++ and executes on an independent PC
under the Microsoft Windows NT 4.0 operating system. Communications with the
ROTSE-III data acquisition system is via RS-232 serial line.

The telescope mount has a maximum slew acceleration of $16.4{^\circ}$/s$^2$ along the right ascension axis, and $20.6{^\circ}/{\rm s}^2$ along the declination axis. Both axes have a maximum slew velocity of $35.0{^\circ}$/s. At these speeds, the mount can slew from horizon to horizon in just over 8 seconds.  A typical alert slew time from the standby (zenith) position takes less than 4 seconds.

During regular system startup the telescope redetermines its absolute home position. This is accomplished by moving the right ascension and declination axes to their limits, where limit switches on the mount are triggered. In this way, our typical pointing error on the sky is up to 25 arcminutes, due to offsets in the home position. To correct for this offset, a test image is taken every night during twilight and automatically calibrated. The true pointing offset is then fed back into the telescope system and the home position is updated. Combining the improved home position with a pointing model calculated by Tpoint \citep{wallace98}, we can achieve rms pointing errors of $\sim 1$ arcminute.

\section{Telescope Enclosure}

The four currently funded ROTSE-III telescope systems are scheduled for
installation in Coonabarabran, Australia, Mt. Gamsberg, Namibia, Bakirlitepe,
Turkey, and Fort Davis, Texas, as listed in Table~\ref{tab:observ}. The
diversity of these locations demands a simple and robust enclosure. To keep
installation efforts to a minimum, the complete enclosure is assembled from
only four pieces: an 8' $\times$ 12' skid, a 90" diameter vertical cylindrical
enclosure, a 24" diameter telescope pier and a motorized hatch cover. With the
exception of the aluminum hatch cover, the entire structure is welded steel.
The total weight of the enclosure is 6750 pounds. By using a steel platform as
the support base, the requisite ground preparation is limited to pouring five
25 cm diameter concrete piers. For locations where wind gusts are not a
problem, the foundation can be as simple as wooden blocks laid directly on the
earth. Since the components are bolted together, the total assembly time in the
field is less than four hours. D \& R Tank Co. of Albuquerque, New Mexico has
been responsible for the development and fabrication of these units. A
photograph of a fully assembled enclosure is shown in Figure~\ref{fig:three}.

The enclosure hatch cover is one of the most critical elements of the
design. For rapid triggered observations of GRBs, access to the full sky is
essential. The cover must swing completely away from the telescope so that
2$\pi$ steradians are visible at all times. Concern about icing problems with
roll-off roofs convinced us to choose a swing-away design as less likely to
fail under adverse weather conditions. For reliability and cost, a Duff-Norton
24" linear actuator with 2400-pound drive force was selected for the
electromechanical drive. This presented an interesting mechanical problem: with
only linear motion, the applied torque must rotate the hatch cover assembly by
180$^\circ$. Fortunately, we found an efficient angle-multiplying crank
mechanism that provided nearly constant torque over the entire travel. The
telescope declination axis is located 4.15" above the highest visual
obstruction, setting the minimum elevation angle at less than 10$^\circ$. Since
the optical tube assembly can clear the hatch aperture in any orientation, the
hatch cover can be closed at any time without fear of damaging the telescope.

The enclosure also houses the three PC-type computers responsible for the operation of the ROTSE-III telescope system.  The primary computer runs the data acquisition software system (see Section 8) and interfaces with the outside world via an ethernet connection to a fiber optic link in the enclosure.  There is also a monitor and keyboard interface to this computer for an on-site operator to manually test and control the system.  The control computer acts as a firewall for the second computer, which is dedicated to running the camera and saving images (see Section 4).  These two computers use the Red Hat distribution of the Linux operating system.  The third machine's sole function is to run the mount control code, compiled under the Microsoft Windows NT 4.0 operating system (see Section 5).  It receives commands from the control computer via a RS-232 serial line.

The electrical system for the enclosure provides power for the data acquisition computers, mount drive motors, hatch cover linear actuator, air blowers, an electric space heater and number of small instruments related to monitoring the weather and the night sky. The absolute peak load is about 3 KVA with the space heater turned on and the linear actuator in motion but the average load is no more than a few hundred watts. Since observatory environments are frequently plagued by electrical storms, the electrical system is designed to protect the equipment from line surges by including a surge protector on the incoming single-phase line and passing the power through an uninterruptable power supply (UPS). The UPS capacity must be at least sufficient to close the hatch cover on detection of power loss. Standard line voltages differ by a factor of two between the United States and virtually everywhere else. We selected most of the electrical components to be compatible with the standards for its destined location and tried to keep the number of items requiring 115 VAC to a minimum. Motor control of the hatch cover is mediated by a custom I/O interface.

The thermal control of the enclosure is quite simple. We decided that air conditioning was far too unreliable for such a remote system. Consequently, the only temperature regulation consists of a thermostat and two 3 CFM Greenheck air blowers (CW-80-L). On exceeding a settable high temperature threshold, both fans turn on and pull air through a large air filter in the enclosure door. This keeps the system from reaching temperatures greater than the ambient external air. To keep the heat load low, the exterior surfaces are painted with Tnemec semi-gloss white polyamide epoxy paint (Pota-Pox series 20, WH01 white). This coating has an 88.2\% solar reflectivity, considerably reducing the daytime heating load. At night, the hatch is fully open so convective flow easily maintains equilibrium.

\section{Environmental Sensing \& Protection}

The key to an autonomous robotic system is the suite of environmental sensors
that inhibit observations under adverse conditions and initiate system shutdown
when such situations arise. The most obvious dangers are high wind and
rain. The Davis Weather Station II provides information on temperature,
humidity, barometric pressure and wind velocity. The temperature data is one of
the required inputs for the telescope focusing model. To avoid damage,
instantaneous wind velocities in excess of 30 mph force automatic system shutdowns. The dew point is determined from the humidity and temperature. Precipitation is sensed with a Vaisala DRD11A Rain Detector and a dual-channel night sky monitor provides information about the average sky brightness in the I and B+V photometric bands. The most likely system failure is the loss of control when a computer `hangs'. To safeguard the system, a watchdog timer circuit must be updated once every second to avoid a shutdown sequence that will close the hatch cover.

\section{Automated Data Acquisition Software}

The ROTSE-III system runs unattended with a fully automated data acquisition (``daq'') system. The daq system runs on a Pentium-III PC with RedHat Linux, and is based on the software developed for the ROTSE-I automated telescope system. It consists of a set of daemons that communicate via shared memory, as in Figure~\ref{fig:four}. The central ``rotse daemon'' ({\tt rotsed}) handles communication between the various daemons. The peripheral daemons each interface with a different aspect of the telescope system.

The clamshell daemon ({\tt clamd}) interfaces with the clamshell control lines. It issues commands to open or close the clamshell. The spot daemon pings the watchdog module in the i/o box every second. If the watchdog module loses contact with the control computer, the system is assumed to have shutdown abnormally and the clamshell is automatically closed via a hardware switch.

The camera daemon ({\tt camerad}) communicates with the CCD camera computer over a TCP/IP socket connection over a private LAN network. The camera software is multi-threaded and can write one image to disk while reading the next image from the CCD camera. The hard disk drives on the camera computer are cross-mounted on the control computer for easy data access.

The weather daemon ({\tt weathd}) communicates with the Davis Weather Station
II and the Vaisala Precipitation Detector. The weather station is
multithreaded, and continuously monitors weather quality. Upon detection of bad weather (when either precipitation in measured or weather statistics exceed configurable limits) the weather daemon issues a Linux signal and the clamshell will be commanded to close until the weather clears.

The Schier mount daemon ({\tt schierd}) communicates with the mount controller computer and focus motor controller via serial lines. The mount daemon implements a Tpoint \citep{wallace98} pointing model for pointing and tracking. It also implements a simple polynomial focus model based on outside temperature and field elevation. The mount daemon monitors the pointing quality from automatically calibrated images to ensure the pointing accuracy on the sky. In this way we achieve our required pointing accuracy of $\sim 1'$.

The astronomical scheduler daemon ({\tt astrod}) schedules observations, system
startup and system shutdown. The astronomical scheduler uses a prioritized
queue scheduler that decides in real-time which configured field is the best to
image next. Gamma-ray burst alerts are automatically put in the front of the queue for immediate processing. Regularly scheduled observations consist of two primary modes of operation. First, a large area of sky is tiled into ``sky patrol'' fields for wide-field transient searches. Second, specific fields can be targeted for long- or short-term monitoring of specific astronomical objects. The standard imaging sequence consists of two 60~s exposures dithered by ${\sim}10$ pixels to relocate bad pixels in consecutive frames. Each field is re-imaged after a configurable cadence interval.

The alert daemon ({\tt alertd}) communicates with the Gamma-ray Burst Coordinates Network (GCN) via a TCP/IP socket interface. The alert daemon also operates a separate server for simulated GCN triggers sent from the University of Michigan and other ROTSE telescopes. Upon receipt of a trigger packet, the alert daemon issues a signal that interrupts the current schedule, stops the mount, aborts the current exposure, and notifies the astronomical scheduler daemon. The total time taken from the receipt of an alert to the start of the first image is typically 5-10~s. The response sequence itself is handled by the astronomical scheduler daemon.

Upon receipt of a prompt burst alert from the GCN, the scheduler initiates an imaging sequence consisting of 10 5~s, 10 20~s, and 80~60s exposures. Under the current configuration, there is a time delay of ${{\sim}7}$~s between exposures due to camera readout and mount dithering. After the prompt burst response, follow-up images of the burst-field are scheduled at logarithmically increasing time intervals. For photometric calibration, each burst response sequence is followed immediately by imaging a high-elevation field of Landolt standard stars \citep{landolt92}.

Manual telescope operation and realtime status monitoring are accomplished with
the Rotse User Shell ({\tt rush}), which is a simple telnet compatible shell
based on {\tt rc}, the configurable open-source shell written by Byron Rakitzis. In addition, selected status variables are sent over a TCP/IP socket to our webserver at {\tt www.rotse.net}, where near real-time status information is available to collaborators and the general public.

\section{Conclusion}
The ROTSE-III project is an example of an optical system that is completely
contrary in spirit to the many recent efforts to build extremely large aperture
instruments. However, it fulfills a significant need in a relatively unexplored
area of astronomy, the search for fast optical transients. The prototype
ROTSE-III instrument is still being tested and debugged at the Los Alamos
National Laboratory in New Mexico but it has already reached a satisfactory
performance that will permit shipment to the Siding Springs Observatory in
Australia by September 2002. The remaining three units are in production and
will be completed by November 2002. We hope our development efforts will serve
as a useful guide to others who wish to pursue similar paths.

\section{Acknowledgments}

The authors are extremely grateful to Dr.~Anna Moore at the Anglo-Australian
Telescope for performing finite element calculations of the primary mirror
gravitational sag.  This work has been supported by NASA grants NAG5-5281 and
F006794, NSF grants AST-0119685 and 0105221, the Michigan Space Grant Consortium, the
Australian Research Council, the University of New South Wales, and the University
of Michigan. Work performed at LANL is supported by NASA SR\&T through
Department of Energy (DOE) contract W-7405-ENG-36 and through internal LDRD
funding. Work at LLNL was performed under the auspices of the DOE,
National Nuclear Security Administration by the University of California,
LLNL under contract No. W-7405-ENG-48.

\clearpage

\begin{deluxetable}{lllllll}
\tablewidth{0pt}
\tabletypesize{\scriptsize}
\tablecaption{Schott Dispersion Constants for Construction Design Run
No.~061900AC\label{tab:one}}
\tablehead{
\multicolumn{7}{c}{Standard Schott Dispersion Formula. Wavelength ($\lambda$)
in microns}\\
\multicolumn{7}{c}{$n^2 = A_0 + A_1 \lambda^2 + A_2 \lambda^{-2} + A_3 \lambda^{-4} + A_4 \lambda^{-6} + A_5 \lambda^{-8}$}\\
\colhead{Element} & \colhead{$A_0$} & \colhead{$A_1$} & \colhead{$A_2$} &
\colhead{$A_3$} & \colhead{$A_4$} & \colhead{$A_5$}}
\startdata
BAL35Y\tablenotemark{a}& 2.4877209 &-1.0549777$\times10^{-2}$ &1.3533860$\times10^{-2}$ &2.2649402$\times10^{-4}$ &-3.9684646$\times10^{-6}$ &3.9817740$\times10^{-7}$ \\
S-FPL51Y\tablenotemark{b}& 2.2186213 &-5.2918817$\times10^{-3}$ &8.4809846$\times10^{-3}$
&8.5916156$\times10^{-5}$ &2.2839583$\times10^{-7}$ &6.9567434$\times10^{-8}$ \\
BK7\tablenotemark{c}&2.2718929 &-1.0108077$\times10^{-2}$ &1.0592509$\times10^{-2}$
&2.0816965$\times10^{-4}$ &-7.6472538$\times10^{-6}$ &4.9240991$\times10^{-7}$ \\
FSIL10C\tablenotemark{d}&2.1042083 &-9.5300149$\times10^{-3}$ &8.6011230$\times10^{-3}$
&1.2304088$\times10^{-4}$ &-1.9274338$\times10^{-6}$ &1.1329291$\times10^{-7}$ \\
\enddata
\tablenotetext{a}{BAL35Y (589613) @ 10.0$^\circ$ C from Ohara glass melt data}
\tablenotetext{b}{S-FPL51Y (497813) @ 10.0$^\circ$ C from Ohara glass melt data}
\tablenotetext{c}{BK7 (517642) Schott standard values}
\tablenotetext{d}{FSIL10C (458678) @ 10$^\circ$ C (fused silica)}
\end{deluxetable}
\clearpage

\begin{deluxetable}{lrrrrl}
\tabletypesize{\scriptsize}
\tablewidth{0pt}
\tablecaption{Construction Design Run No.~061900AC at $T=+10^\circ$ C for
ROTSE-III Telescopes}
\tablehead{
\colhead{Description} & \colhead{Radius of} & \colhead{Axial
Thickness} & \colhead{Min. Clear} & \colhead{Conic Constant} &
\colhead{Material}\\
\colhead{} & \colhead{Curvature (mm)} & \colhead{(mm)} & \colhead{Aperture (mm)} & \colhead{} & \colhead{}}
\startdata
primary mirror   & -1620.000 & \nodata  & 453.0 & -1 & Pyrex\\
                 &           & -449.000 &       &    & air\\
secondary mirror & $\infty$  & \nodata  & 227.0 &  0 & Pyrex\\
                 &           &  176.00  &       &    & air\\
lens \#1         &    83.850 &   12.000 & 125.0 &  0 & BAL35Y\\
                 &    92.594 &   50.692 & 118.0 &  0 & air\\
lens \#2         &   145.136 &    6.000 &  91.0 &  0 & BAL35Y\\
                 &    61.304 &   29.337 &  82.0 &  0 & air\\
lens \#3         &  -651.890 &    7.000 &  76.0 &  0 & BAL35Y\\
                 &  -354.010 &   34.082 &  76.0 &  0 & air\\
filter simulator & $\infty$  &    4.000 &  68.0 &  0 & BK7\\
                 & $\infty$  &   10.000 &  66.0 &  0 & air\\
shutter plane    &           &    0.000 &  63.0 &    &    \\
                 &           &   10.000 &  63.0 &    & air\\
lens \#4         &    82.435 &   10.000 &  60.0 &  0 & S-FPL51Y\\
                 & $\infty$  &   11.945 &  58.0 &  0 & air\\
cryostat window  & $\infty$  &    4.762 &  49.0 &  0 & FSIL10C\\
                 & $\infty$  &    7.737 &  47.0 &  0 & vacuum\\
CCD detector     & $\infty$  &  \nodata &  39.1 &  0 & \\
\enddata
\label{tab:two}
\end{deluxetable}

\clearpage

\begin{deluxetable}{llll}
\tablewidth{0pt}
\tablecaption{Locations of ROTSE-III Observatories\label{tab:observ}}
\tablehead{
\colhead{Site} & \colhead{Longitude} & \colhead{Latitude} &
\colhead{Altitude (m)}}
\startdata
Coonabarabran & $149^\circ 3' 40.3''$ E & $31^\circ 16' 24.1''$ S
& 1149 \\
Mt. Gamsberg & $16^\circ 30' 00''$ E & $23^\circ 16' 18''$ S & 1800 \\
Bakirlitepe & $30^\circ 20' 0''$ E & $36^\circ 49' 30''$ N & 2550 \\
Fort Davis & $104^\circ 1' 20.1''$ W & $30^\circ 40' 17.7''$ N & 2075 \\
\enddata
\end{deluxetable}

\clearpage

\begin{figure}
\plotone{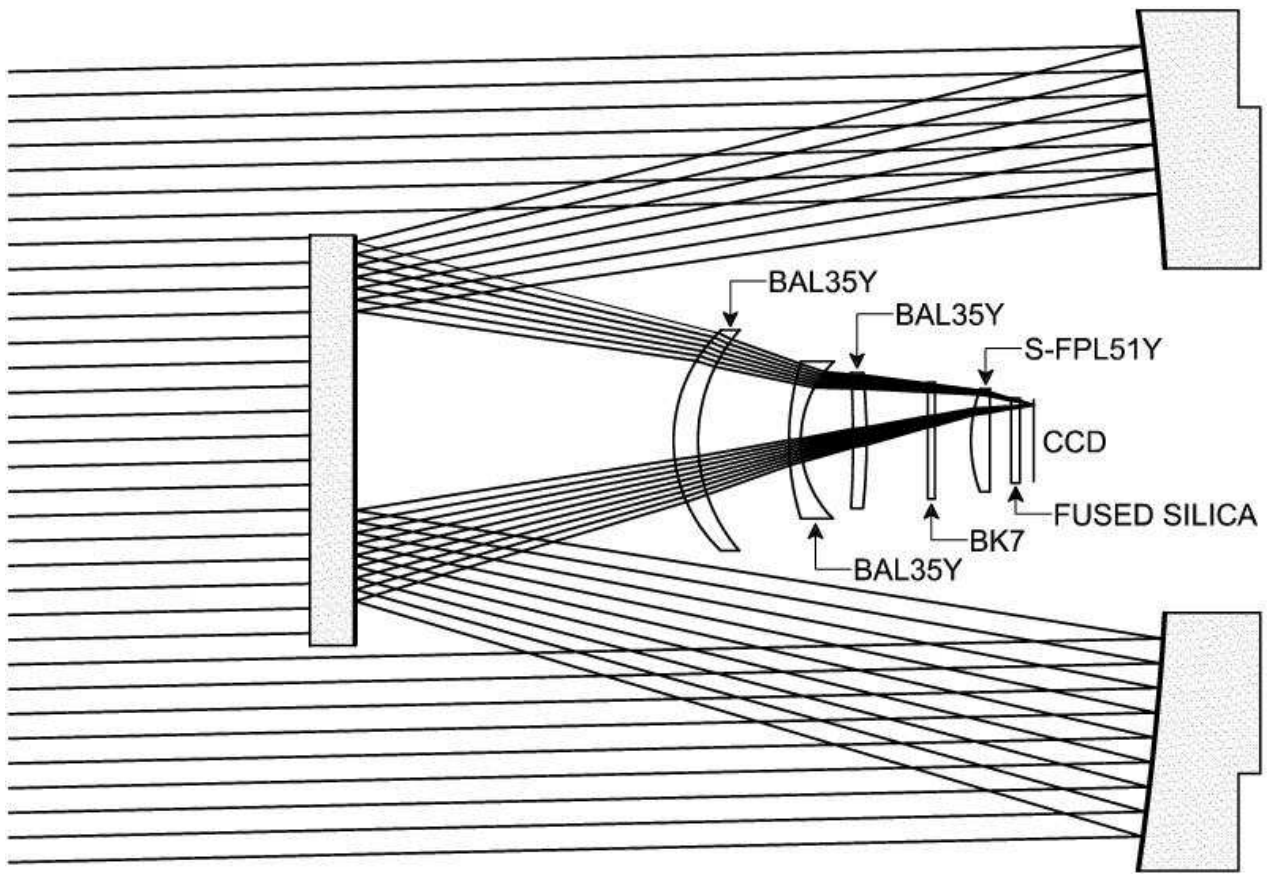}
\caption{\label{fig:one}The ROTSE-III telescope optical design.  Rays are shown for the extreme of the $2{^\circ}.64$ field-of-view.}
\end{figure}

\clearpage

\begin{figure}
\plotone{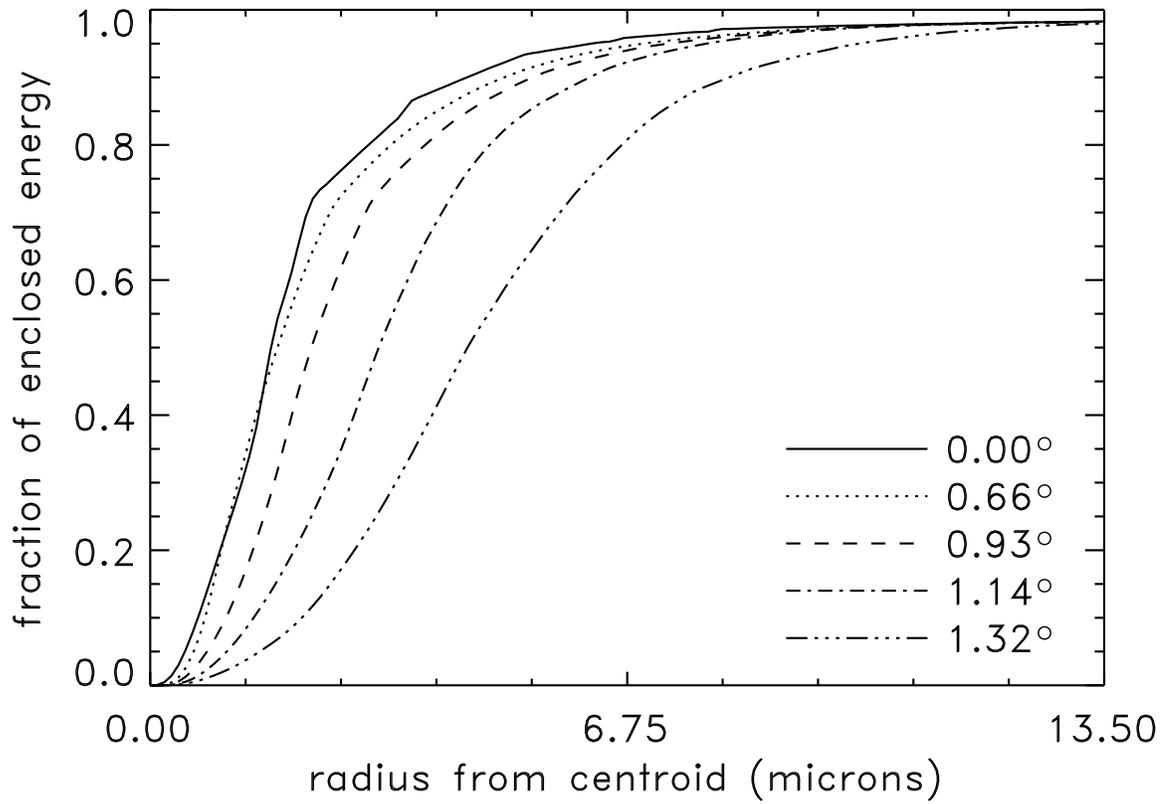}
\caption{\label{fig:two}The polychromatic fractional encircled energy for the ROTSE-III optical design.  The curves show the performance on-axis and at 39\%, 78\%, 96\% and 100\% of the CCD field-of-view.}
\end{figure}

\clearpage

\begin{figure}
\plotone{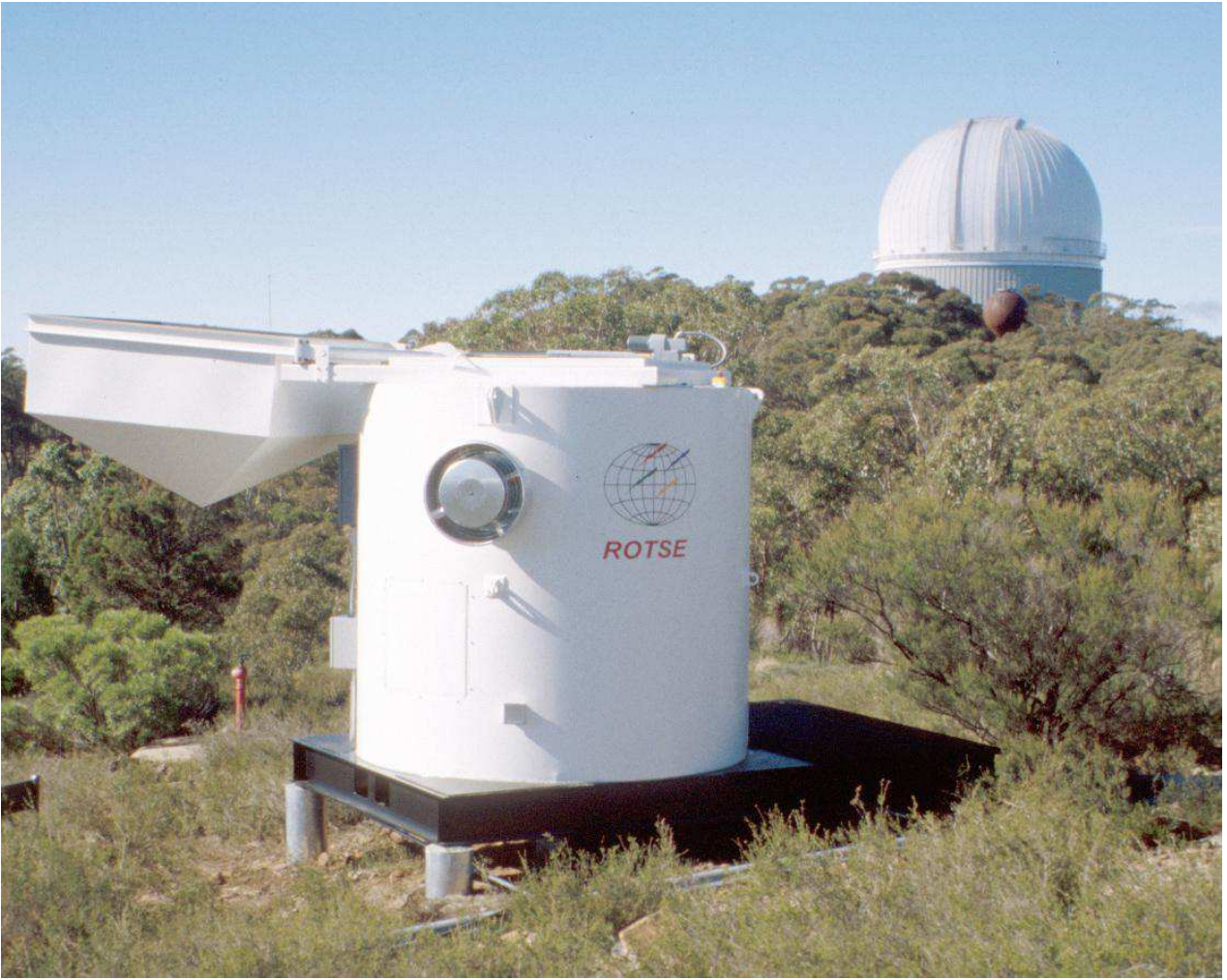}
\caption{\label{fig:three}The ROTSE-III telescope enclosure at the Siding Springs Observatory near Coonabarabran, Australia with the hatch cover fully open.  The telescope optics have not yet been installed.  The 4-meter Anglo-Australian telescope is in the background.}
\end{figure}

\clearpage

\begin{figure}
\plotone{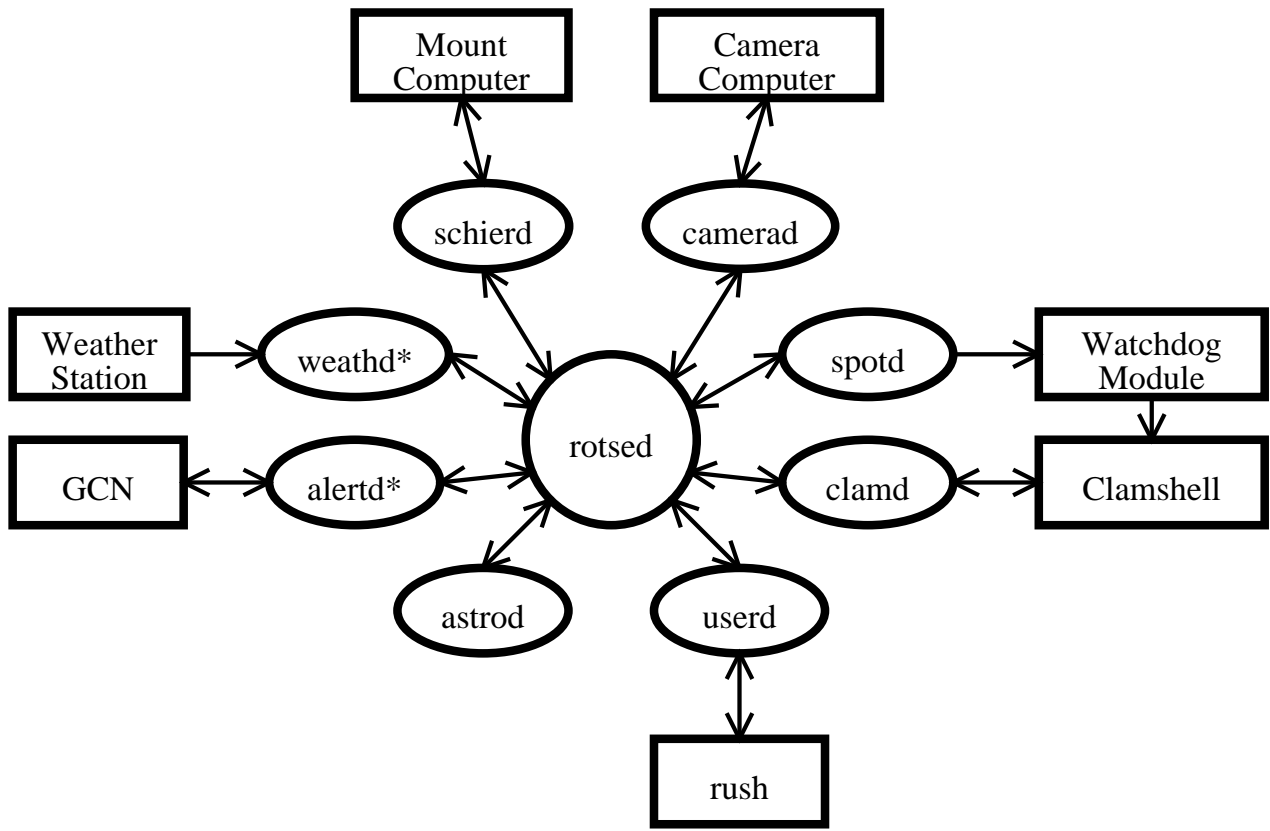}
\caption{\label{fig:four}The ROTSE-III daq system is made up of a number
of interconnected daemons which communicate through the central ``rotse
daemon'' ({\tt rotsed}). The astronomical scheduler daemon is denoted {\tt
astrod}. Each daemon interfaces with a different logically distinct aspect of the telescope system. The daemons denoted with asterisks are also able to send Linux signals to interrupt the camera, mount, and scheduler.}
\end{figure}


\begin{thebibliography}{99}

\bibitem[Akerlof et al. 1999]{akerlof99} Akerlof, C., et al. 1999, \nat, 398, 400

\bibitem[Akerlof et al. 2000]{akerlof00} Akerlof, C., et al. 2000, \apjl, 532, 25L

\bibitem[Barthelmy 1998]{sdb97} Barthelmy, S., 1998, AIP Conference Proceedings,
428, 99

\bibitem[Beatty 1982]{beatty82} Beatty, J., 1982, \skytel, 63, 469

\bibitem[Groot et al. 1997]{groot97} Groot, P., et al., 1997, \iaucirc 6584

\bibitem[Jeas 1981]{jeas81} Jeas, W. C., 1981,
Military-Electronics/Countermeasures, 7, 47

\bibitem[Kehoe et al. 2001]{kehoe01} Kehoe, R., et al., 2001, \apjl, 554, 159L

\bibitem[Kehoe et al. 2002]{kehoe02} Kehoe, R., et al., 2002, \apj, 577, 845

\bibitem[Landolt 1992]{landolt92} Landolt, A., 1992, \aj, 104, 340

\bibitem[Lee 1997]{lee97} Lee, B., 1997, \apjl, 482, 125L

\bibitem[Park et al. 1997-a]{park97a} Park, H-S., et al., 1997-a, \apj, 490, 99

\bibitem[Park et al. 1997-b]{park97b} Park, H-S., et al., 1997-b, \apjl, 490, 21L

\bibitem[Pravdo 1999]{pravdo99} Pravdo, S., 1999, \aj, 117, 1616

\bibitem[Sampson 1913-a]{sampson13a} Sampson, R., 1913-a, Philosophical
Transactions of the Royal Society of London, 213, 27

\bibitem[Sampson 1913-b]{sampson13b} Sampson, R., 1913-b, \mnras, 73, 524

\bibitem[Stokes 2000]{stokes00} Stokes, G., Evans, J., Viggh, H., Shelly, F.,
\& Pearce, E., 2000, Icarus, 148, 21

\bibitem[Wallace 1998]{wallace98} Wallace, P. T., 1998, www.tpsoft.demon.co.uk/pointing.htm.

\bibitem[Williams et al. 1999]{williams99} Williams, G., et al., 1999, \apjl, 519, 25L

\bibitem[Wynne 1949]{wynne49} Wynne, C. G., 1949, Proceedings of the Physical
Society (London), 62, 772

\bibitem[Wynne 1967]{wynne67} Wynne, C. G., 1967, \ao, 6, 1227

\end{thebibliography}
\end{document}